\documentclass[11pt,a4paper]{article}
\usepackage[utf8]{inputenc}
\def\parfrac#1#2{{\left(\frac{#1}{#2}\right) }} 
\usepackage{caption}
\usepackage{subcaption}
\usepackage{jcappub}
\usepackage{bm}
\usepackage{epsfig}
\usepackage{bbold}
\usepackage{graphicx}
\usepackage{amsmath}
\usepackage{amssymb}
\usepackage{amsbsy}
\usepackage{color}
\usepackage{slashed}
\usepackage{afterpage}
\usepackage{psfrag}
\usepackage[noabbrev]{cleveref}
\usepackage{dsfont}
\usepackage{enumerate}
\usepackage[shortlabels]{enumitem}
\usepackage[utf8]{inputenc}
\usepackage{amsmath}
\usepackage[dvipsnames]{xcolor}
\usepackage{graphicx}
\usepackage{braket}
\usepackage{physics}
\usepackage{siunitx}
\usepackage{float}
\usepackage{tikz-feynman}
\tikzset{snake/.style={decorate, decoration=snake}}

\title{\begin{center}
   Superradiant axion clouds around asteroid-mass primordial black holes
   \end{center}}

\author[a]{Nuno P. Branco,}
\author[b]{Ricardo Z. Ferreira,}
\author[a]{João G. Rosa}

\affiliation[a]{Univ Coimbra, Faculdade de Ci\^encias e Tecnologia da Universidade de Coimbra and CFisUC, Rua Larga, 3004-516 Coimbra, Portugal}
\affiliation[b]{Institut de Física d’Altes Energies (IFAE) and Barcelona Institute of Science and Technology (BIST), Campus UAB, 08193 Bellaterra, Barcelona, Spain}

\emailAdd{popebranco@hotmail.com}
\emailAdd{rzambujal@ifae.es}
\emailAdd{jgrosa@uc.pt}

\makeatletter
\gdef\@fpheader{}
\makeatother

\abstract{We analyze the dynamics and observational signatures of axion clouds formed via the superradiant instability around primordial black holes, focusing on the mass range $10^{14}-10^{18}$ kg where the latter may account for all the dark matter. 
We take into account the leading effects of axion self-interactions, showing that, even though these limit the number of axions produced within each cloud, 
a large number of superradiant axions become free of the black hole’s gravitational potential and accumulate in the intergalactic medium or even in the host galaxy, depending on their escape velocity.  This means that primordial black hole dark matter may lead to a sizeable astrophysical population of non-relativistic axions, with masses ranging from 0.1 eV to 1 MeV, depending on the primordial black hole mass and spin.  
We then show that if such axions couple to photons their contribution to the galactic and extragalactic background flux, mainly in the X-ray and gamma-ray band of the spectrum, is already beyond current observational limits for a large range of parameters that are, therefore, excluded. We finish by showing the prospects of the Athena X-ray telescope to further probe this co-existence of primordial black holes and axions.}

\begin{document}

\maketitle

\section{Introduction}\label{sec: Introduction}

Superradiant scattering in classical systems was originally considered by Zel'dovich, who showed that scalar waves can be amplified upon scattering off a rotating absorbing cylinder, provided that their frequency satisfies the superradiance condition $\omega<m\Omega$, where $m$ is the azimuthal angular momentum quantum
number and $\Omega$ is the cylinder’s angular velocity \cite{zel1972pis}. 

Rotational superradiance is, in fact, characteristic of any rotating absorbing medium, in particular Kerr black holes \cite{Starobinsky:1973aij,StarobinskyChurilov, Teukolsky:1973ha, Press:1973zz, Teukolsky:1974yv}. Over four decades ago, Press and Teukolsky envisaged the idea of a ``black hole bomb" \cite{Press:1972zz, Cardoso:2004nk} or superradiant instability, which would result from multiple superradiant scatterings induced by enclosing the black hole with a reflecting mirror. Although this may {\it a priori} seem like a purely academic exercise, a natural mirror-like effect occurs for massive fields, which are confined in quasi-bound states by the black hole's gravitational potential \cite{Damour:1976kh, Zouros:1979iw, Detweiler:1980uk, Furuhashi:2004jk, Cardoso:2005vk, Dolan:2007mj, Rosa:2009ei, Dolan:2012yt, East:2017ovw, East:2017mrj, Dolan:2018dqv}. If these states satisfy the superradiance condition, their occupation numbers grow exponentially fast (presumably from small initial quantum fluctuations), leading to a cloud of particles surrounding the black hole. This provides an extremely efficient particle production mechanism, which is fueled by the black hole's rotational energy, therefore depleting the latter's spin down to the angular velocity that saturates the superradiance condition, i.e. for which $\omega=m\Omega$.

Since the black hole's gravitional potential is Coulomb-like at large distances, the spectrum of quasi-bound states for a (spinless) field of mass $\mu$ around a black hole of mass $M$ has a Hydrogen-like form:
\begin{equation}
    \label{eq:Hydrogenlikespectrum}
    \hbar\omega_n\simeq \mu c^2\left(1-\frac{\alpha^2}{2n^2}\right)
\end{equation}
in the non-relativistic regime where the dimensionless mass coupling 
\begin{eqnarray}
\alpha=\mu G M/\hbar c \simeq 0.75 \parfrac{M}{M_\odot} \parfrac{\mu}{10^{-10} \text{eV}} \, ,
\end{eqnarray}
with $n$ denoting the principal quantum number and $M_\odot$ the solar mass. To leading order, the superradiance condition may be recast in the form $\alpha < \tilde{a}/2(1+\sqrt{1-\Tilde{a}^2})$, where $0\leq \tilde{a}\leq 1$ is the black hole's dimensionless spin parameter, so that even for extremal black holes the dimensionless mass coupling satisfies $\alpha<1/2$ and the spectrum is essentially Hydrogen-like. 

Astrophysical black holes suffer from superradiant instabilities in the presence of ultra-light fields, with $\mu\lesssim 10^{-10}$ eV. This has attracted a significant interest in the recent literature as a novel ``laboratory" to probe the existence of such particles, including in particular the QCD axion and other axion-like particles\footnote{In this work we will generically refer to these particles as axions.} \cite{Arvanitaki:2009fg, Arvanitaki:2010sy, Rosa:2012uz, Witek:2012tr, Brito:2014wla, Arvanitaki:2016qwi,  Brito:2017wnc, Rosa:2017ury, Baumann:2018vus,   Ikeda:2018nhb, Berti:2019wnn, Baumann:2019eav, Cardoso:2018tly, Edwards:2019tzf,Mehta:2020kwu,Blas:2020nbs}, as well as hidden photons or massive gravitons \cite{Rosa:2011my, Pani:2012bp, Brito:2013wya,Baryakhtar:2017ngi,Caputo:2021efm}.

The universe may, however, be (or have been) filled with a large number of much lighter black holes, potentially as light as the Planck mass $M_P\simeq2.18\times10^{-8}$ \si{kg}. These primordial black holes (PBHs), originally proposed by Hawking \cite{Hawking:1971ei, Carr:1974nx}, may have formed in the early universe through a plethora of possible different mechanisms (see e.g. \cite{Byrnes:2021jka,Green_2021, Carr:2016drx} for recent reviews), and can account for a fraction of the present dark matter density. Sub-solar mass PBHs are prone to superradiant instabilities for heavier fields. These can include Standard Model particles such as pions \cite{Ferraz:2020zgi} but potentially many other novel particles predicted in extensions of the Standard Model, including axions and other dark particles \cite{Rosa:2017ury, March-Russell:2022zll}, whose mass does not necessarily lie in the above-mentioned ultra-light range. Moreover, superradiant instabilities of sub-solar mass PBHs may occur on cosmological timescales even if PBHs are born with low spin parameters as we will explicitly discuss here (see also \cite{March-Russell:2022zll}).

Although the PBH cosmological abundance is quite constrained for a wide range of masses \cite{Carr:2020gox,Green_2021}, PBHs can, in fact, still account for all the dark matter in the present universe if they lie in the asteroid-mass range $10^{14}\lesssim M\lesssim 10^{18}$ kg \cite{Green_2021} where they can trigger superradiant instabilities of axions with masses in the eV-MeV range.
In this work, we study the dynamics of axion superradiant instabilities around PBHs in these mass ranges. We will show that a significant cosmological axion population may result from PBH superradiance if the PBHs account for all the dark matter and such a population can give rise to detectable observational signatures if the axion couples to photons.

In our dynamical study, we will take into account the leading effects of axion self-interactions that, albeit suppressed by tiny coupling constants, are greatly enhanced by the exponentially large numbers of axions that constitute the superradiant clouds, as shown in \cite{Gruzinov:2016hcq, Baryakhtar:2020gao} (see also \cite{Yoshino:2012kn, Yoshino:2015nsa, Omiya:2020vji, Omiya:2022gwu, Omiya:2022mwv}). We will see that, as first pointed out in \cite{Baryakhtar:2020gao}, self-interactions may quench the growth of the particle number within the clouds, but that this is achieved at the expense of a large number of axions becoming free of the PBH's gravitational attraction. These ``ionized'' axions may be confined within the PBH's host galaxy, if released with sufficiently small velocity ($\alpha < 10^{-2}$ as we will see below), or otherwise travel through the intergalactic medium. Hence, even in scenarios where only a small fraction of the axions remains bound to the clouds, PBH superradiance is nevertheless a very efficient axion production mechanism.

A key parameter in our analysis will, of course, be the PBH natal spin. On the one hand, when PBHs are born from the gravitational collapse of large overdensities once they become sub-horizon in a radiation-dominated epoch, typical spin parameters are expected to lie at or even below the percent level, $\tilde{a}\lesssim 0.01$ \cite{Chiba:2017rvs, DeLuca:2019buf, Mirbabayi_2020}. If, on the other hand, PBH formation occurs in an early matter-dominated era, 
the low ambient pressure favours an anisotropic collapse and the consequent generation of near-extremal PBHs \cite{Harada_2017}. While there may be several other formation mechanisms yielding intermediate spin parameters, we will consider these two limiting regimes to illustrate how the PBH spin affects superradiant axion production.

The last ingredient of our study is the possibility that the produced axions decay in photon pairs such that PBH superradiance can then contribute, for example, to the galactic or extragalactic background flux, distort the CMB or even overheat some galaxies.
We will use existing limits on the amount of dark matter that can decay into photons \cite{Ng:2019gch,Ballesteros:2019exr,Bolliet:2020ofj,Laha:2020ivk,Foster:2021ngm,Wadekar:2021qae} to constrain the axion-photon coupling in the different mass ranges, and we will also briefly discuss how future X-ray telescopes such as Athena \cite{Neronov:2015kca} will further test the co-existence of axions and dark matter PBHs. 

This work is organized as follows. In the next section we discuss the axion mass range for which superradiant production is efficient and compute the maximum number of axions produced. In Section \ref{sec: Clouddynamics} we analyze in detail the dynamics of the superradiant clouds, taking into account the PBH's mass and spin depletion, the effect of axion self-interactions and the axion decay into photons. In particular, we analytically compute the number of axions generated within the superradiant clouds and the number of axions that get ionized, and confirm our results with numerical simulations. In Section \ref{sec: Electromagnetic signatures of superradiant axion clouds}, we discuss the electromagnetic signatures of PBH-axion superradiance associated with the decay of axions into photons and use observational data (mainly of galactic and extragalactic background fluxes) to place constraints on the axion-photon coupling constant. Finally, we summarize our results and discuss future prospects in Section \ref{sec: Conclusion}.

\section{Conditions for efficient superradiance}
We start with a short overview of black hole superradiance, discussing the necessary conditions for it to occur and determining in which cases the superradiant cloud can grow significantly within the lifetime of the universe. This will allow us to identify the range of masses, of both the PBHs and the axions, for which superradiance can play an important role.

\begin{figure}[t]
    \centering
    \includegraphics[width=10cm]{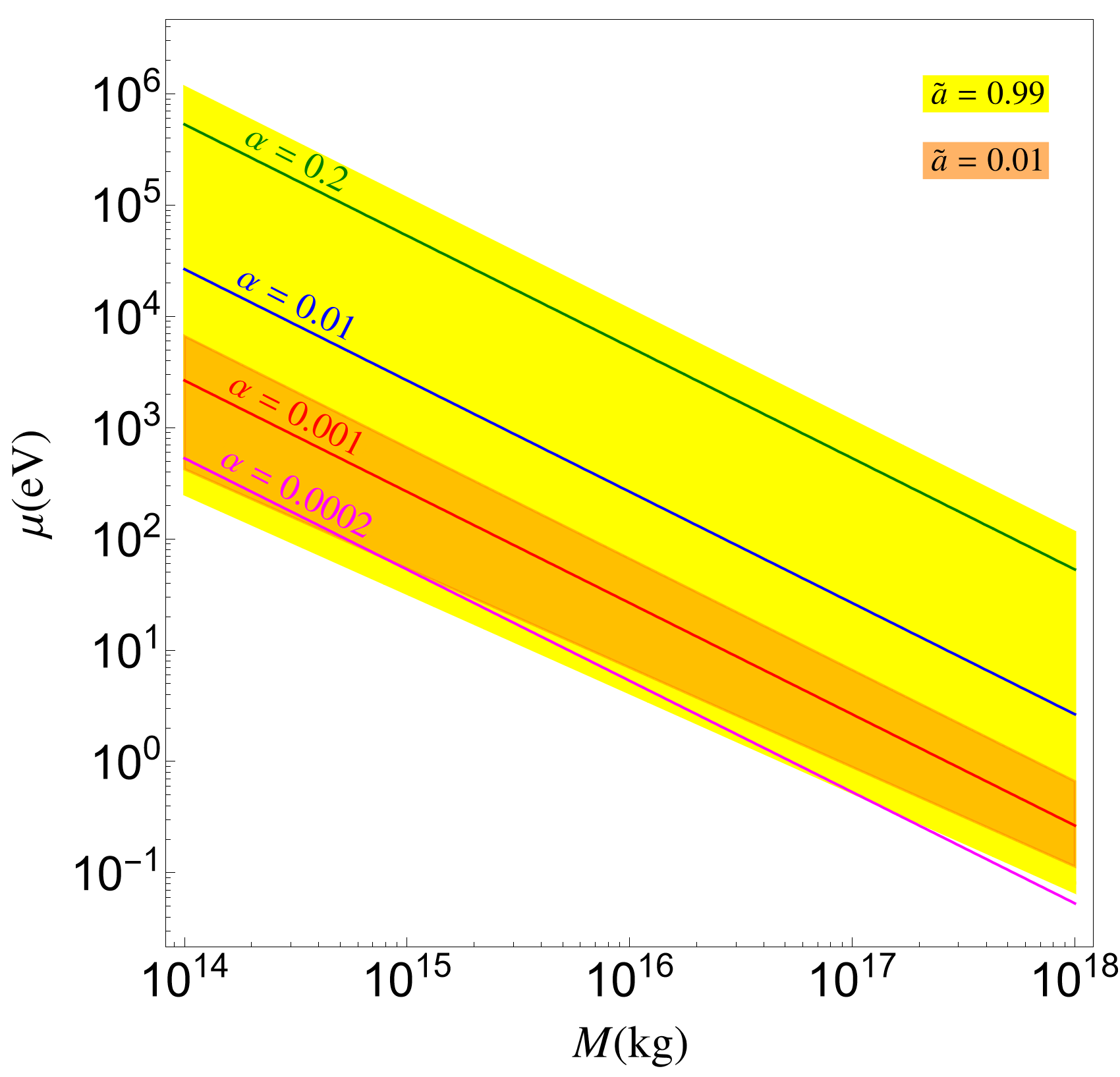}
    \caption{Yellow and orange bands are the regions of PBH mass $M$ and axion mass $\mu$ where the superradiant condition is satisfied and the cloud grows to its maximal value (in the absence of interactions) within the lifetime of the universe for two representative values of the initial black hole spin: $\tilde{a}=0.01$ (yellow band), and $\tilde{a}=0.99$ (orange band). 
Solid lines are iso-countours of $\alpha$. 
}
\label{fig: Parameter space for the ALP.}
\end{figure}

As mentioned above, whilst the superradiance condition is satisfied, numerous axions are copiously produced in quasi-bound states in the vicinity of the black hole at the expense of its rotational energy, $N \sim \Delta J /m$. The time scale for the superradiant modes to grow depends strongly on the angular momentum number $l$ of the quasi-bound state and decreases with $\alpha$ as $\alpha^{-4l-5}$. Therefore, the fastest growing state has quantum numbers $\left|nlm\right>=\left|211\right>$ and is populated much faster than all others. The number of axions in this level grows as $N_2 = \exp(\Gamma_2\, t)$ where $\Gamma_2\equiv \Gamma_{211}$ is the the superradiant rate given, for small $\tilde{a}$, by  \cite{Detweiler:1980uk,Cardoso:2004nk} \footnote{For large values of $\tilde{a}$ and $\alpha$ the analytical expression for the superradiant rates in eq. \ref{eq: superrate} is no longer accurate. A phenomenological expression for $\Gamma_{211}$ that provides a good fit in this regime is given by \cite{Ferraz:2020zgi}
\begin{equation}
\label{eq: improvedsuperradiantrate}
\Gamma_{211}^\text{imp}=\Gamma_{211}\left[1+\frac{3}{128} \frac{\tilde{a}^{9/5}}{\sqrt{1-\tilde{a}}}\left(\frac{r_+ \mu}{ \tilde{a}}\right)^6\right]\,.
\end{equation}
We will use this corrected expression in this work whenever $\tilde{a}$ and $\alpha$ are large.}
\begin{equation}
\label{eq: superrate}
    \Gamma_{nlm} = 2r_+ C_{nlm} (m\Omega- \omega_{nlm})\alpha^{4l+4}\mu   
\end{equation}
where the coefficient $C_{nlm}$ is given by
\begin{eqnarray}
    C_{nlm}=\frac{2^{4l+1}(n+l)!}{n^{2l+4}(n-l-1)!} \prod_{k=1}^{l}\left[k^2(1-\tilde{a})^2-\left(\tilde{a}m-2 \tilde{r}_+ \alpha \right)^2\right]
\end{eqnarray}
and the radial coordinate of the inner (Cauchy) and outer (event) horizons are given by $r_\pm=GM \tilde{r}_+ = G M(1\pm\sqrt{1-\tilde{a}^2})$, with the angular velocity of the black hole at the event horizon being given by $\Omega=\Tilde{a}/(2r_+)$. For slow-rotation $\tilde{a}\ll 1$ and small $\alpha$ the rate above simplifies to  $\Gamma_2  \simeq  \alpha^8\mu (\tilde{a}-4\alpha)/24$.

The number of particles in a given state grows until the superradiance condition is saturated, i.e. $\omega=\Omega$. In the absence of interactions, this happens for the dominant mode ($\left|211\right>$) at a time $t_\text{max}=  \ln{(N_2^\text{max})}/\Gamma_2$ when the number of axions in this state is
\begin{equation}
	\label{eq: N max}
	N_2^\text{max}= \Delta J \simeq (\tilde{a}_i-\tilde{a}_f) \,G M^2 \, ,
\end{equation}
where $\tilde{a_i}$ and $\tilde{a}_f=4\alpha$ are, respectively, the initial and final spins of the black hole and we have neglected the variation of the black hole mass, $M_i \simeq M_f$, which is a good approximation for $\alpha \ll 1$  since $\Delta M/M=\alpha\tilde{a}\Delta J/J $.
However, superradiance only plays an important role if the cloud fills up in a time scale smaller than the age of the universe, i.e. 
\begin{eqnarray} \label{eq:superradiance in the age of the universe}
        t_\text{uni} \gtrsim t_\text{max}=  \ln{(N_2^\text{max})}/\Gamma_2\,.
\end{eqnarray}

Figure \ref{fig: Parameter space for the ALP.} shows the range of PBH and axion masses where the superradiance condition and the lifetime condition, eq. \ref{eq:superradiance in the age of the universe}, are simultaneously satisfied for two benchmark values of $\tilde{a}$ ($\tilde{a}=0.01$, $\tilde{a}=0.99$) that we will consider throughout this work and that, as described earlier, are representative of two PBH formation mechanisms: during radiation domination (small spin) and during an early matter domination (near extremal). As also previously discussed, we will restrict our analysis to the mass range $M \sim 10^{14}-10^{18}$ kg for which the PBHs can account for all the dark matter \cite{Green_2021}. The figure shows that superradiance can efficiently produce axions with masses from 0.1 eV to a few keV for slowly spinning PBHs, while in the near-extremal limit this range is widened up to MeV masses.

\section{Cloud dynamics with self-interactions}
\label{sec: Clouddynamics}

In the absence of interactions the number of particles within the superradiant cloud grows until it reaches its maximal value $N\simeq N_2^\text{max}$. However, if the bosonic field has self-interactions, as in the case of axions, this growth can be quenched significantly. The goal of this section is to study the role of self-interactions in the evolution of the cloud.

We consider the typical axion Lagrangian\footnote{Additional couplings of the axion to other Standard Model fields can be considered but do not affect our results because we will be consider axions with masses below the electron mass that can only decay into photons.} 
\begin{eqnarray} \label{eq: Axion Lagrangian}
	{\cal L}_a\supset \frac{1}{2} \partial_\mu a \, \partial^\mu a + \mu^2 f_a^2 \left[1-\cos\left(\frac{a}{f_a} \right) \right]+ c_{a\gamma \gamma} \frac{\alpha_\text{EM}}{8\pi} \frac{a}{f_a} F_{\mu \nu} \tilde{F}^{\mu \nu}  \,,
\end{eqnarray} 
where $a$ denotes the axion field, $\alpha_{EM}$ is the electromagnetic fine structure constant, $f_a$ is the axion decay constant and $c_{a\gamma\gamma}$ a dimensionless coefficient proportional to the electromagnetic anomaly. In this work, we fix the coefficient $c_{a\gamma\gamma} =1$ so that the coupling $f_a$ fixes both the strength of the self-interactions and the coupling to photons, although our results can be easily generalized for other values of this coefficient. For later convenience we also define $g_{a \gamma \gamma}\equiv \alpha_\text{EM}/(2\pi f_a) $. 

On the one hand, the cosine potential induces self-interactions amongst the axions. For $a \ll f_a$ the leading self-interaction term is quartic and characterized by a dimensionless coupling $\lambda= \mu^2/f_a^2\simeq 10^{-32}(\mu/\mathrm{keV})^2(10^{10}\, \mathrm{GeV}/f_a)^{2}$.  Although this is typically very small, processes within a superradiant cloud are Bose-enhanced by huge factors due to the exponentially large number of axions, as we will explore below.
On the other hand, through the axion-photon coupling axions can decay into two photons at a rate
\begin{equation}
	\label{eq: decayrate}
\Gamma_{a\gamma\gamma}= \frac{g_{a\gamma \gamma}^2 \mu^3}{64 \pi} \,.
\end{equation}
Although we will consider parametric regimes where this decay has a negligible influence on the dynamics of the clouds, it will be crucial when exploring the observational signatures of PBH-axion superradiance.

We will begin by considering the dynamical effects of self-interactions amongst axions in the main superradiant bound states. As we will see, this may lead to a significant number of axions escaping the cloud.

\subsection{Axions within the superradiant cloud}
\label{sec: Boltzmann equations}

Self-interactions become progressively more important as $a$ approaches $f_a$. The growth of the cloud is halted roughly when the quartic self-interactions start to compete with the gravitational binding energy, as analyzed in detail in  \cite{Baryakhtar:2020gao}.

\begin{figure}[t]
	\centering
	\begin{tikzpicture}
		\node[draw] at (-1.2, 1.5)    {211};
		\node[draw] at (-1.2, -1.5)    {211};
		\node[draw] at (1.2, -1.5)    {322};
		\node[draw] at (1.2, 1.5)    {BH};
		\draw[gray, thick] (-1,1) -- (1,-1);
		\draw[gray, thick] (-1,-1) -- (1,1);
		\filldraw[black] (0,0) circle (2pt);
		\node[draw] at (3.8, 1.5)    {322};
		\node[draw] at (3.8, -1.5)    {322};
		\node[draw] at (6.2, -1.5)    {211};
		\node[draw] at (6.2, 1.5)    {$\infty$};
		\draw[gray, thick] (4,1) -- (6,-1);
		\draw[gray, thick] (4,-1) -- (6,1);
		\filldraw[black] (5,0) circle (2pt);
	\end{tikzpicture}
	\caption{\small 
	Leading scattering processes for the dynamics of the superradiant cloud: depletion (left), and replenishment (right). We follow the notation of \cite{Baryakhtar:2020gao} and denote the superradiant states by their quantum numbers $nlm$, the states absorbed by the black hole by BH and those that escape the cloud by $\infty$.}
	\label{fig: Most dominant processes.}
\end{figure}
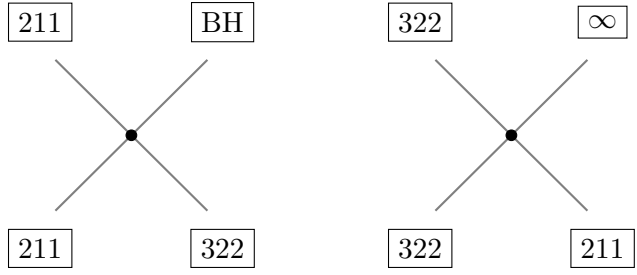

The relevant $2\rightarrow 2$ scattering processes in this dynamics have been identified in \cite{Baryakhtar:2020gao} and involve the superradiant levels $\left| 211 \right>$ and $\left| 322 \right>$, and axions that either escape to infinity $(\infty)$ or are absorbed by the black hole $(\text{BH})$. They can be summarized as
(see also Figure \ref{fig: Most dominant processes.}):
\begin{eqnarray}
&\textit{Depletion:}& \qquad  \left| 211 \right> +  \left| 211 \right> \, \overset{\Gamma_\text{D}\,}{\longrightarrow}  
	\, \text{BH} + \left| 322 \right> 
 \label{Depletion process}
 \\	&\textit{Replenishment:}& \qquad  \left| 322 \right>  + \left| 322 \right>\, \overset{\Gamma_\text{R}\,}{\longrightarrow}   \, \infty + \left| 211 \right>\, ,  \label{Replenishment process}
\end{eqnarray}
 where the nomenclature refers to the effect of these processes on the dominant $\left| 211 \right>$ state, and occur respectively at a rate \cite{Baryakhtar:2020gao}
\begin{eqnarray} \label{eq: Depletion rate}
    \Gamma_\text{D} &=& 4\times 10^{-7} \alpha^7 \lambda^2 h(\tilde{a})\mu \\ 
    \label{eq: Replenishment rate}
        \Gamma_\text{R} &=&  10^{-8} \alpha^4 \lambda^2 \mu\,, 
\end{eqnarray} 
where we have defined $h(\tilde{a})\equiv 1+\sqrt{1-\tilde{a}^2}$. The axions that escape to infinity have angular momentum $m=3$  and energy larger than the axion rest mass, while those absorbed by the black hole have $m=0$, i.e.~are in a non-superradiant bound state. Therefore, the latter changes the black hole mass but not its spin.
To track how the processes above affect the evolution of the cloud we solve the following Boltzmann equations for the number of particles in the $\left|211\right>$ and $\left|322\right>$ states, $N_2$ and $N_3$ respectively,\footnote{We neglect stimulation effects that may enhance the decay into photons  under certain conditions  \cite{Rosa:2017ury}.}
\begin{eqnarray}
              \frac{dN_2}{dt}&=& N_2 \left[ \Gamma_2 -\Gamma_{a\gamma\gamma} + N_3 \left(\Gamma_\text{R}N_3 -2\Gamma_\text{D}N_2 \right) \right]  \label{eq: ODE for N2} \\
       \frac{dN_3}{dt}&=&N_3 \left[ \Gamma_3-\Gamma_{a\gamma\gamma}+N_2 \left(\Gamma_\text{D}N_2-2\Gamma_\text{R}N_3 \right) \right] \label{eq: ODE for N3}\, ,
\end{eqnarray}
where $\Gamma_3 \equiv \Gamma_{322}$ is the superradiance rate of the $\left|322\right>$ state given in eq. \ref{eq: superrate}. The first terms on the right hand side capture the effect of superradiance; the second terms correspond to the decays of the axions in the cloud into photons; the third terms incorporate the effects of depletion and replenishment due to axion self-scatterings.     
 Superradiance drains spin and mass from the black hole by populating the leading superradiant levels but part of the black hole mass is recovered in the depletion process. Therefore, the Boltzmann equations above need to be complemented by the corresponding evolution equations for the black hole mass $M$ and spin $J$
\begin{eqnarray}
       \frac{dM}{dt}&=& -\mu\,(\Gamma_2N_2+\Gamma_3N_3-\Gamma_\text{D}N_2^2N_3)\label{eq: Mass evolution} \\
       \frac{dJ}{dt}&=&-(\Gamma_2N_2+2\Gamma_3N_3) \label{eq: Spin evolution} \,.
\end{eqnarray}

We now proceed to describe the dynamics in more detail. We will neglect the terms in $\Gamma_3$ and $\Gamma_{a \gamma \gamma}$ in the analysis as they are sub-leading in the range of parameters that we consider. However, we explicitly include them in our numerical analysis of the above system of equations. The decay rate into photons will be crucial in Section \ref{sec: Electromagnetic signatures of superradiant axion clouds} when studying the observational signatures of this scenario. We note that there are other sub-leading scattering processes, as well as gravitational wave emission, that we are not including but can be found in \cite{Baryakhtar:2020gao}.

Initially, the $\left|211\right>$ level starts to grow exponentially fast through the superradiance term, $\Gamma_2$. As $N_2$ grows, the terms in $N_2^2$ start to move a considerable number of particles to the level $\left|322\right>$. An equilibrium is eventually reached when
\begin{equation}
\label{eq: equi}
    \frac{dN_2}{dt}=\frac{dN_3}{dt}=0
\end{equation}
and the two levels reach the equilibrium values
\begin{eqnarray}
\label{eq: equilibrium value for N2}
N_2^\text{eq}&=&\frac{2}{\Gamma_\text{D}}\sqrt{\frac{\Gamma_\text{R}\Gamma_{2}}{3}} \simeq\frac{50}{\sqrt{3}  } \frac{\sqrt{\tilde{a}-4\alpha}}{
\alpha } \, \parfrac{f_a}{\mu}^2 \,  \\  N_3^\text{eq}&=& \sqrt{\frac{\Gamma_{2}}{3\Gamma_\text{R}}} \simeq\frac{2000}{\sqrt{3}} \alpha^2\sqrt{\tilde{a}-4\alpha} \, \parfrac{f_a}{\mu}^2 \,, \label{eq: equilibrium value for N3}
\end{eqnarray}
where in the last equalities we have used eqs. \ref{eq: superrate}, \ref{eq: Depletion rate} and \ref{eq: Replenishment rate} in the slow-rotation limit. 

There are two main conditions that determine the fate of the cloud: 1) whether the black hole loses a significant part of its spin within the life-time of the universe; 2) whether the equilibrium value is larger or smaller than the maximum number of particles that can be produced via superradiance (eq. \ref{eq: N max}). These conditions are connected with the strength of self-interactions, controlled by $f_a$, and lead to different regimes in the evolution of the cloud that we describe below and summarize in Table \ref{fig: Table with regimes}. 

\begin{table}
\caption{Different regimes for the evolution of the superradiant cloud. \label{fig: Table with regimes}}
\vspace{0.1cm}
\begin{tikzpicture}
\centering
\node (table) [inner sep=0pt] {
\begin{tabular}{c||c|c|c|c}
       \textbf{Spin-down} & \textbf{No} & \textbf{Incomplete}  & \textbf{Complete} & \textbf{Fast}  \\
    \hline \hline
 \textbf{Regime} 
 &$f_a \lesssim f_a^{\text{spin-}\downarrow} \,$  &  $f_a^{\text{spin-}\downarrow} \lesssim f_a \lesssim f_a^\text{no-spin} \,$  & $f_a^\text{no-spin} \lesssim f_a \lesssim f_a^\text{no-eq}\,$   & $f_a > f_a^\text{no-eq} \,$  \\
     \hline \hline
            \textbf{Equation} & \ref{eq: f no spin-down} & \ref{eq: f no spin-down}, \ref{eq: f complete spin-down} & \ref{eq: f complete spin-down}, \ref{eq: f fast spin-down} & \ref{eq: f fast spin-down}\\
    \hline 
\end{tabular}
};
\draw [rounded corners=.5em] (table.north west) rectangle (table.south east);
\end{tikzpicture}
\label{table with regimes}
\end{table}

\textit{\underline{No spin-down}:}
When the interactions are relatively strong (smaller $f_a$), the cloud reaches the equilibrium state more quickly, with a smaller equilibrium number of axions $N_2^\text{eq}$. Therefore, the amount of spin extracted from the black hole during the lifetime of the universe is rather small and $N_2$ remains at the equilibrium value. We call this the \textit{no spin-down} regime. 
The emission of axions to infinity is also less efficient because both $N_2$ and $N_3$ are relatively small.

To estimate the range of parameters that lie within this regime we compare the time scale $t_J$ for the black hole to lose a significant amount of its spin 
\begin{eqnarray}
    t_J \simeq \frac{J}{\dot{J}}\simeq\frac{\tilde{a}G M^2}{N_2^\text{eq}\Gamma_2}
\end{eqnarray}
with the age of the universe. In terms of $f_a$, \textit{no spin-down} corresponds to
\begin{eqnarray} \label{eq: f no spin-down}
  f_a<f_a^{\text{spin-}\downarrow} \equiv 4.4\times10^9\frac{\tilde{a}^{1/2}
}{(\tilde{a}-4\alpha)^{3/4}}\left(\frac{\si{keV}}{\mu}\right)^{1/2}\left(\frac{4\times10^{-4}}{\alpha}\right)^{5/2}\si{GeV} \, .
\end{eqnarray} 

\textit{\underline{Incomplete spin-down}:}
If $f_a>f_a^{\text{spin-}\downarrow}$ the equilibrium numbers are larger than in the previous regime and, as consequence, $t_J<t_\text{uni}$ and the PBH spin starts decreasing significantly, therefore considerably reducing the superradiance growth rate $\Gamma_2$.  Nevertheless, as long as this rate is sufficiently large to compensate for the cloud's depletion due to self-interactions  ($\Gamma_2 N_2 \gg \Gamma_D N_2^2 N_3, \Gamma_R N_3^2 N_2$), an equilibrium configuration is approximately maintained, with $N_2\simeq N_2^{\mathrm{eq}}(t)$ as given by eq.~\ref{eq: equilibrium value for N2}
with a decreasing $\Gamma_2(t)$. 
 In this adiabatic regime, we may obtain a solution for $N_2$ by substituting $N_2\simeq N_2^{\mathrm{eq}}(t)$ in the spin evolution eq.~\ref{eq: Spin evolution} and solving for $\tilde{a}$ (taking into account that $\Gamma_3N_3\ll \Gamma_2N_2$ and neglecting the PBH mass change). An analytical solution can then be obtained for slowly rotating PBHs away from the superradiance threshold, $\tilde{a}\gtrsim 4\alpha$ for which $\Gamma_2\propto \tilde{a}$, yielding: 
\begin{eqnarray} \label{eq: incomplete spin-down}
          N_2(t>t_J)=\frac{N_2^\text{eq}}{1+\frac 1 2 \frac{N_2^\text{eq}}{N_2^\text{max}}\Gamma_2 (t-t_{J})}\, , 
\end{eqnarray}
where $N_2^{\mathrm{eq}}$ refers to the equilibrium value attained before the PBH begins its spin-down at $t_J$. Note that within the age of the universe the present occupation number of the $|211\rangle$ state is only significantly reduced compared to its initial equilibrium value if $N_2^\text{eq}/N_2^\text{max}\gtrsim (\Gamma_2 t_\text{uni})^{-1}$.

\textit{\underline{Complete spin-down}:}
For even larger values of $f_a$, and so also of $N_2^\text{eq}$, the adiabatic regime is efficient enough to spin-down the BH so much that superradiance is no longer effective in populating the cloud and the dynamics becomes instead dominated by the depletion and replenishment processes in eqs. \ref{Depletion process} and \ref{Replenishment process}.
 Once the PBH spins down sufficiently to nearly saturate the superradiance condition at $t_\text{no-spin}$, we may neglect the superradiance terms in the evolution equations for $N_2$ and $N_3$ in eqs. \ref{eq: ODE for N2} and \ref{eq: ODE for N3}. The resulting coupled system of equations admits solutions of the form $N_3/N_2 =  \Gamma_D/(2\Gamma_R) \equiv d \approx \mathrm{const.}$, such that $N_2$ evolves in time as
\begin{eqnarray}
    \label{eq: complete spin-down}
N_2(t>t_\text{no-spin}>t_\text{J})=\frac{N_{\text{no-spin}}}{\left(1+3d\, \Gamma_\text{D} N^2_{\text{no-spin}}(t-t_{\text{no-spin}})\right)^{1/2}}
\end{eqnarray}
where $N_\text{no-spin}$ is the value of $N_2$ at the time $t_\text{no-spin}$. We may estimate $t_\text{no-spin}$ by matching eqs. \ref{eq: incomplete spin-down} and \ref{eq: complete spin-down} yielding
\begin{equation}    t_{\text{no-spin}}\simeq6.8\times10^{12}\left(\frac{\si{keV}}{\mu}\right)\left(\frac{10^{11}\si{GeV}}{f_a}\right)^4\left(\frac{4\times10^{-4}}{\alpha}\right)^2 \si{years} \,.  \label{eq: t no spin}
\end{equation}
\textit{Complete spin-down} will occur whenever $t_{\text{no-spin}}<t_{\text{uni}}$, which corresponds to decay constants
\begin{equation}  \label{eq: f complete spin-down}
f_a>f_a^\text{no-spin} \equiv 4.7\times10^{11}\left(\frac{\si{keV}}{\mu}\right)^{1/4}\left(\frac{4\times10^{-4}}{\alpha}\right)^{1/2}\si{GeV}\,.
\end{equation}

\begin{figure}[t]
    \centering
    \includegraphics[width=11.7cm]{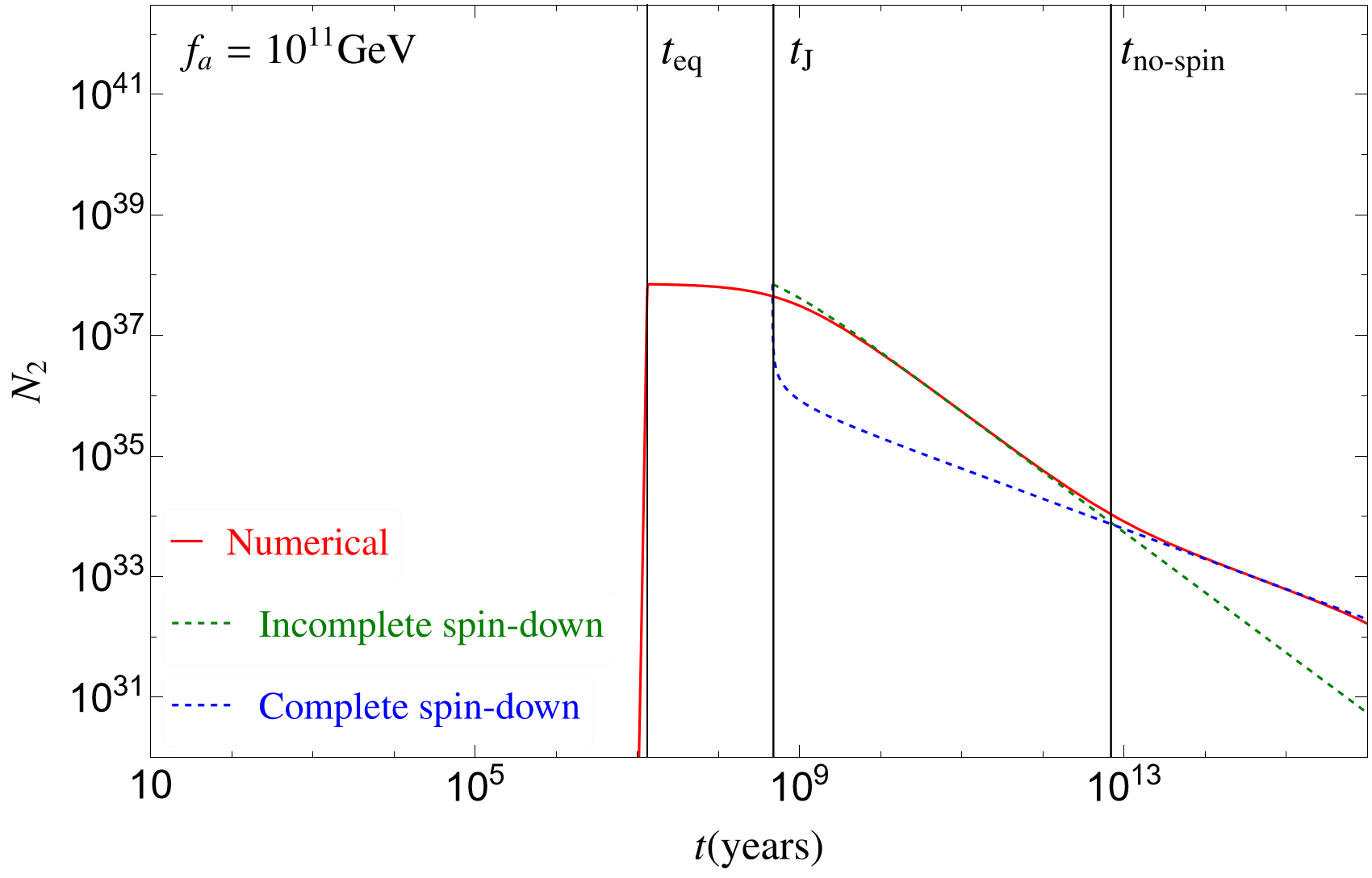}
    \caption{\small  Numerical evolution (solid red curve) of the number of particles in the dominant $|211\rangle$ state for a PBH with $M=10^{14}$ kg and $\tilde{a}=0.01$ and an axion with $\mu= 1$ keV and $f_a=10^{11}$ GeV. The dashed green and blue lines correspond to the analytical solutions in the \textit{incomplete spin-down} and \textit{complete spin-down} regimes given in eqs. \ref{eq: incomplete spin-down} and \ref{eq: complete spin-down}. Note that although $f_a=10^{11}$ GeV corresponds to the \textit{incomplete spin-down} regime, here we can also see the \textit{complete spin-down} phase because we have extended the time domain beyond the age of the universe. }
    \label{fig: matching.}
\end{figure}

In Figure \ref{fig: matching.} we show the numerical evolution of $N_2$ for $f_a=10^{11}$ GeV where the three different dynamical stages discussed so far are clearly illustrated: (1) superradiant growth until $t_\text{eq}$, at which the equilibrium is attained; (2) at $t_J$ the PBH starts spinning down and the equilibrium value decreases adiabatically until (3) $t_\text{no-spin}$, after which superradiance shuts down and the dynamics of the cloud is controlled by the self-interactions. We leave a more general discussion of the numerical results to Section \ref{sec: Numerics}, although we note the clear agreement between our analytical description and the numerical solution in this figure.

\textit{\underline{Fast spin-down}:}
The final regime occurs for even larger $f_a$ when $N_\text{eq}$ is so large that the cloud saturates at $N_2^\text{max}$ \footnote{We neglect here the subsequent growth of the cloud through the sub-leading superradiant levels as these rates are small and do not affect the dynamics.}, i.e. superradiance shuts down before an equilibrium is reached. This happens when $N_2^\text{eq}> N_2^\text{max}$ or equivalently for
\begin{eqnarray} \label{eq: f fast spin-down}
f_a>f_a^\text{no-eq} \equiv  1.8\times10^{13}\left(\frac{\alpha}{4\times10^{-4}}\right)^{3/2}
(\tilde{a}-4\alpha)^{1/4} \, \si{GeV}\, .
\end{eqnarray}
With the superradiance source turned off, the dynamics is again controlled by self-interactions just like in the \textit{complete spin-down regime}. Therefore, $N_2$ will still evolve like in eq. \ref{eq: complete spin-down} but with $N_\text{no-spin},t_\text{no-spin}$ replaced by $N_2^\text{max}$ and $t_{\max}$.
Noteworthy, this is the only regime among those we described where most of the axions are still in the cloud today and only a small fraction has escaped. However, as we will show in Section \ref{sec: Electromagnetic signatures of superradiant axion clouds}, when discussing the observational impact of the dynamics, the \textit{fast spin-down} is not yet accessible with current data since the associated axion-photon coupling is too small while the total number of axions produced remains roughly unchanged.

\subsection{Axions ionized from the cloud}
So far we have described the dynamics of the axions within the superradiant cloud and we have seen that self-interactions can strongly quench its growth.
However, the replenishment processes in eq. \ref{Replenishment process} (and Figure \ref{fig: Most dominant processes.}) are continuously producing axions that escape the cloud. At the end of the dynamics most of the axions are, in fact, outside the cloud (except in the \textit{fast spin-down} regime where self-interactions play a negligible role). At the same time, the coupling to photons allows these particles to decay into photon pairs at a rate $\Gamma_{a \gamma \gamma}$ given in eq. \ref{eq: decayrate}. The number of particles that escape the cloud obeys therefore
\begin{equation}
\label{eq: ninnodecay}
    \frac{dN_\infty}{dt} = \Gamma_\text{R}N_3^2N_2 - \Gamma_{a \gamma \gamma} N_\infty
\end{equation}
being fully determined by the solutions for $N_2$ and $N_3$ derived in Section \ref{sec: Boltzmann equations}. 

What we have left to understand is the fate of these ionized axions, in particular, if they become bounded to the PBH host galaxy and how many of them decay to photons.
Let us begin by addressing the first point. Conservation of energy in the replenishment process implies that the ionized axions are emitted with non-relativistic energy $E_\infty =\alpha^2\mu/ 72$ and a velocity
\begin{eqnarray}
    v_\infty &\simeq& \frac \alpha 6 = 6.7\times10^{-5} \parfrac{\alpha}{4\times 10^{-4}} \, .
\end{eqnarray}
This value needs to be compared with the escape velocity of the host galaxy. Axions with $v_\infty<v_\text{esc}$ will be bound to the host galaxy while the remaining ones will escape to the intergalactic space. This distinction will be important in the next section when comparing the flux of photons from superradiant axion decay with the observational data. In the next section we will use X-ray and gamma-ray data from the Milky Way (MW), Andromeda (M31) and Segue 1 (Seg1) galaxies, and bounds on the rate of energy injection in the Leo-T galaxy. Their escape velocities are respectively given by  $v_\text{esc}^\text{MW}=0.0018,v_\text{esc}^\text{M31}=0.0016,v_\text{esc}^\text{Seg1}=0.0002$  \cite{EscapeVelocityMilkyWay,Kafle:2018amm,Baushev:2012ke} \footnote{The escape velocity increases as one approaches the galactic center. Here, we conservatively use $v_\text{esc}$ evaluated at a kpc distance from the galactic center (except for Segue 1 where the quoted value is estimated from the gravitational potential at the center of the galaxy). For Leo-T we take $v_\text{esc}^\text{Leo-T}={\cal O}(10^{-4})$ as a benchmark value \cite{Faerman:2013pmm}.}.  
Hence, we conclude that the ionized axions will typically remain bound to the host galaxy for slowly rotating black holes with $\tilde{a}\lesssim 0.01$, since the superradiance condition requires in this case $\alpha \lesssim 10^{-3}$.

\begin{figure}[t]

\begin{subfigure}{.494\linewidth}
  \includegraphics[width=\linewidth]{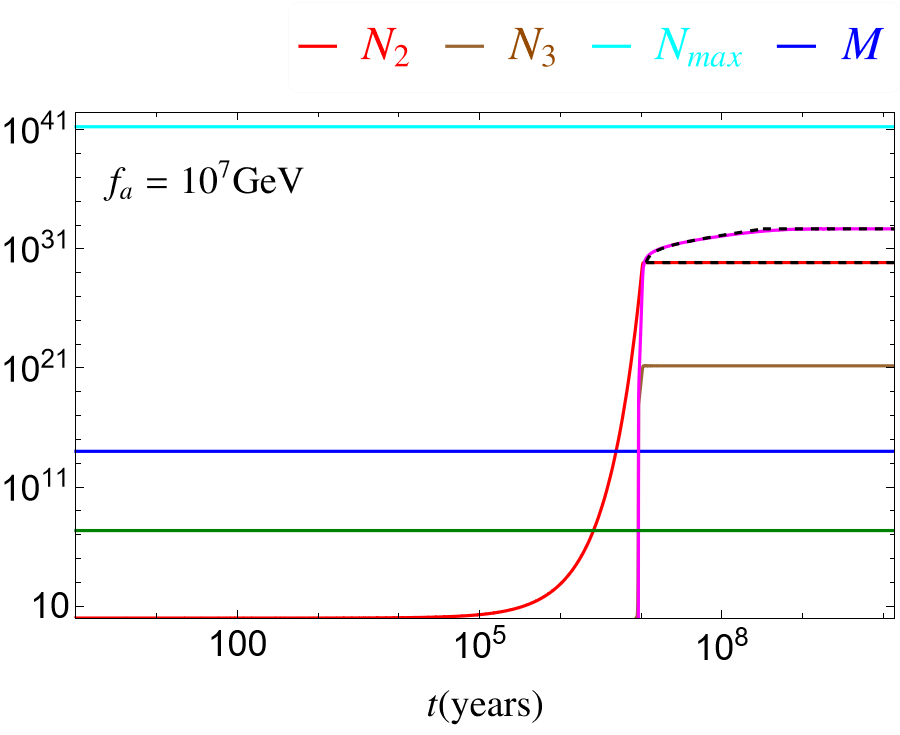}
\end{subfigure}\hfill 
\begin{subfigure}{.494\linewidth}
  \includegraphics[width=\linewidth]{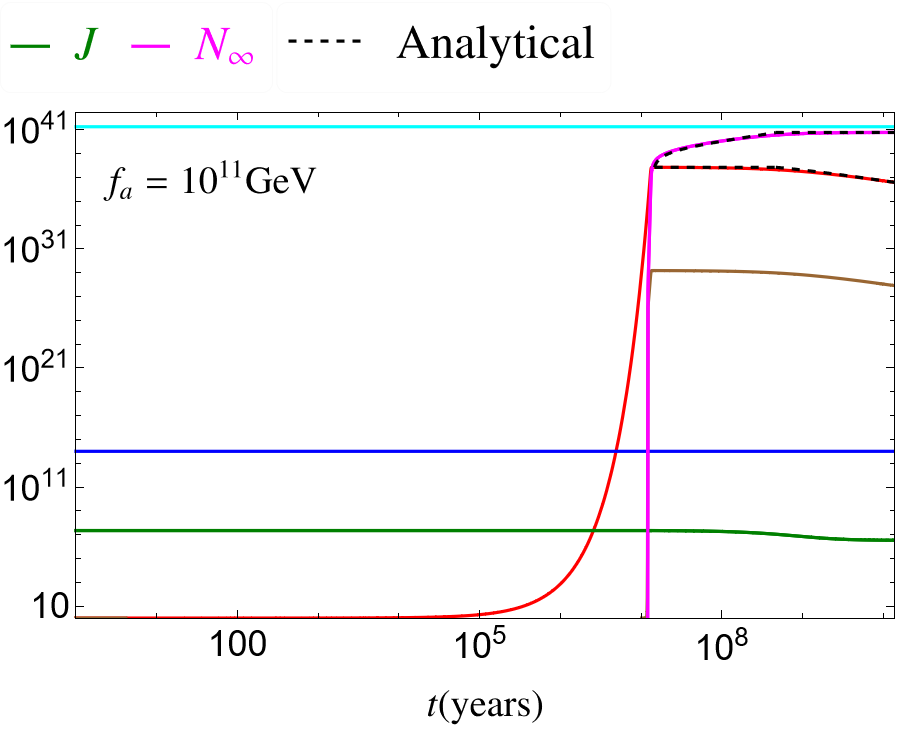}
\end{subfigure}

\medskip 
\begin{subfigure}{.494\linewidth}
  \includegraphics[width=\linewidth]{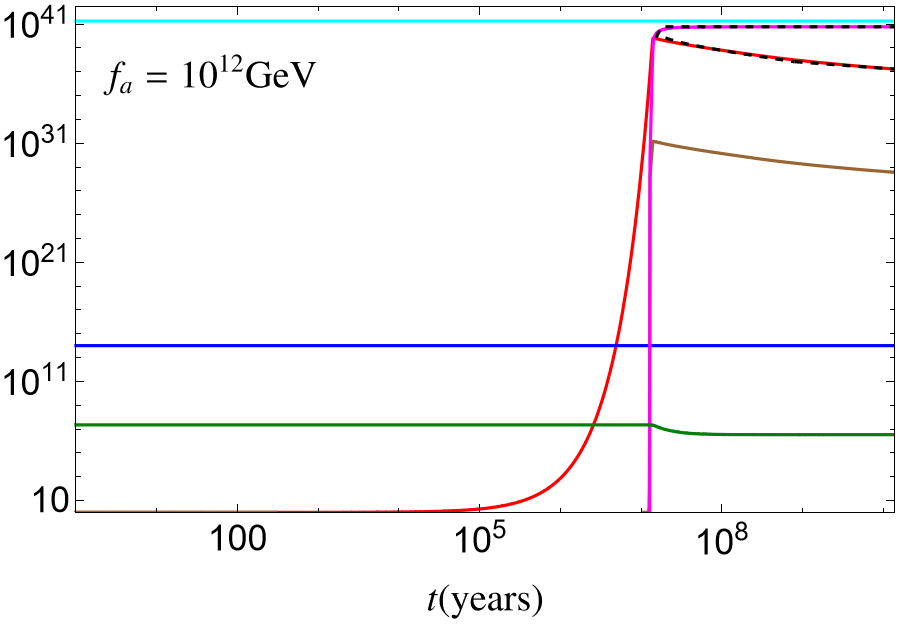}
\end{subfigure}\hfill 
\begin{subfigure}{.494\linewidth}
  \includegraphics[width=\linewidth]{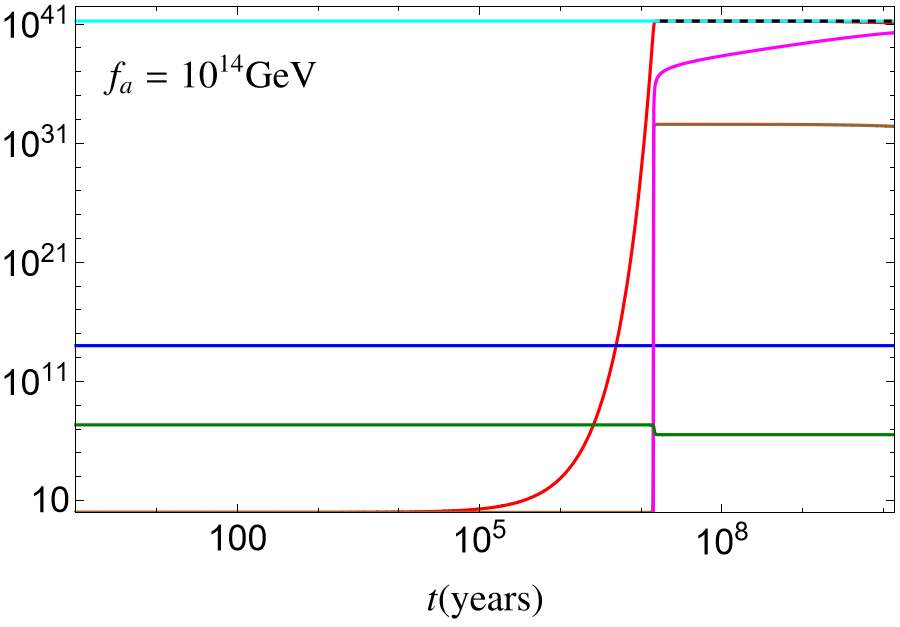}
\end{subfigure}

\caption{\small Numerical solutions for $\mu=1$ keV, $M=10^{14}$ kg, $\Tilde{a}=0.01$ and different $f_a$ values. In red/brown we plot the number of axions in the dominant/sub-dominant superradiant state ($N_2/N_3$) and in magenta the number of axions that escape the black hole's gravitational potential ($N_\infty$). In blue/green we show the variation of the black hole's mass ($M$)/angular momentum ($J$). In cyan we plot the maximum allowed axion number ($N_2^\text{max}$). Analytical solutions for $N_2$ and $N_\infty$ are plotted in dashed black between $t_{\text{eq}}\leq t \leq t_{\text{uni}}$ where they overlap with the corresponding numerical solutions.}
\label{fig: Numerical solutions.}
\end{figure}

Secondly, as the number of free axions grows, it is possible that at some point their decay rate into photons, i.e.~the first term on the right-hand side of eq. \ref{eq: ninnodecay}, balances the rate at which they are produced via the replenishment processes. This will happen when 
\begin{eqnarray} \label{eq_inf}
    N_\infty \Gamma_{a \gamma \gamma} = \Gamma_R N_3^2 N_2 \,.
\end{eqnarray}
For the range of parameters that we studied in this work, this balance only occurs in the \textit{no-spin down regime} where $N_2, N_3$ reached the (constant) equilibrium values in eqs. \ref{eq: equilibrium value for N2} and \ref{eq: equilibrium value for N3}. Therefore, during this phase $N_\infty$ will grow linearly in time at the rate $\Gamma_R N_3^2 N_2$ until it saturates the condition \ref{eq_inf} at a time 
\begin{equation} \label{eq: equilibrium for Ninf}
t_{\text{decay}}\simeq 3.1\times10^{10}\left(\frac{f_a}{10^{8}\si{GeV}}\right)^2\left(\frac{\si{keV}}{\mu}\right)^3\si{years} \,
\end{equation}
when 
\begin{equation}
\label{eq: decayrestriction}
    N_\infty= N_\infty^\text{eq} \equiv 1.9\times10^{40} \frac{(\tilde{a}-4\alpha)^{3/2}}{h(\tilde{a})}\left(\frac{\alpha}{4\times10^{-4}}\right)^7\left(\frac{f_a}{10^{8}\si{GeV}}\right)^4\left(\frac{\si{keV}}{\mu}\right)^4 \,.
\end{equation}
As we can see from eq. \ref{eq: equilibrium for Ninf}, only for small values of the axion decay constant $f_a$,
\begin{eqnarray} \label{eq: f where decay to photons is relevant}
    f_a \lesssim  6.5 \times 10^7 \parfrac{\mu}{\text{keV}}^{3/2} \text{GeV} \,
\end{eqnarray}
is the equilibrium reached within the lifetime of the universe.
The effects of the decays are shown in the upper left corner of Figure \ref{fig: Numerical solutions.} where we can see the linear growth of $N_\infty$ until the equilibrium value $N_\infty^\text{eq}$ is attained.

\subsection{Total number of axions and PBH spin loss}
\label{sec: total number of axions}

The spin of the black hole feeds the cloud with axions. Therefore, in the regimes where the decay into photons is less efficient than the superradiant axion production (all values of $f_a$ except those in eq. \ref{eq: f where decay to photons is relevant} that we discuss below), the final number of axions can be calculated by conservation of angular momentum: 
\begin{eqnarray} \label{eq: Ntot}
N_\text{tot}= N_2 + 2 N_3 + 3N_\infty \simeq    \Delta J
\end{eqnarray}
where we have used the fact that the axions that are emitted to infinity have angular momentum $m=3$. Moreover, in the previous subsection we concluded that the number of axions outside the cloud is larger than the number within the cloud for a broad $f_a$ range (see also Figure \ref{fig: Numerical solutions.}). Therefore, 
$\Delta J(t_\text{uni}) \simeq 3N_\infty(t_\text{uni})$ except for the \textit{fast spin-down} regime where most axions are still bounded to the PBH within the cloud and $\Delta J(t_\text{uni}) \simeq N_2^\text{max}$.

The other exception is for $f_a$ larger than \ref{eq: f where decay to photons is relevant} (i.e.~small couplings). In this case, $N_\infty(t_\text{uni}) \simeq N_\infty^\text{eq} \gg N_2$, but the total number of photons produced is even larger $N_\gamma(t_\text{uni}) = 2\, t_\text{uni} \Gamma_{a\gamma \gamma}  N_\infty^\text{eq} \gg  N_\infty^\text{eq} $ and this is where most of the black hole spin is dumped.

\subsection{Numerical analysis} 
\label{sec: Numerics}

In Figure \ref{fig: Numerical solutions.}
we show numerical solutions of the system of equations comprising the Boltzmann equations \ref{eq: ODE for N2} and \ref{eq: ODE for N3} for the number of axions in the two main superradiant states, and the dynamical equations for the PBH mass and spin, \ref{eq: Mass evolution} and \ref{eq: Spin evolution}. We consider a fixed value of $\alpha=4\times 10^{-4}$ (corresponding to $M=10^{14}$ kg and $\mu=1$ keV) and spin $\tilde{a}=0.01$, and four different values of $f_a$ that are representative of the different regimes identified in Table \ref{table with regimes}. In all cases, the PBH mass remains approximately constant throughout the cosmological evolution until the present day and the cloud reaches an equilibrium at around $ 10^7$ years. 

For low values of $f_a$ ($f_a=10^7$ GeV  upper left corner), the cloud is in the \textit{no spin-down} regime and remains approximately stable until today. However, as we increase $f_a$ to  $f_a=10^{11}$ GeV (upper right corner) the system enters the \textit{incomplete spin-down} regime where the cloud reaches a larger equilibrium number that starts to slowly decrease over time as the PBH loses a sizeable amount of its spin (green line in Figure \ref{fig: Numerical solutions.}).

In the lower left corner we show the evolution in the \textit{complete spin-down} regime with $f_a=10^{12}$ GeV. In this case, the equilibrium number is so large that the PBH loses spin much faster and superradiance eventually shuts down before the present day. Afterwards, the dynamics of the cloud is dominated by axion self-interactions.

 Finally the lower right corner shows the evolution for $f_a=10^{14}$ GeV, within the \textit{fast spin-down} regime, where the cloud reaches the maximal occupation number, at which superradiance shuts down, before reaching the equilibrium state. In this regime, self-interactions are rather weak so $N_2$ remains roughly constant and most of the axions are presently still within the cloud rather than escaping the latter. 

 We end this section by noting that, even though in some cases the axion's lifetime is smaller than the age of the universe, the fact that the superradiant rate $\Gamma_2 > \Gamma_{a\gamma \gamma}$ enforces that the total number of axions is always growing over time. The effect of the decay into photon pairs is only significant in the upper left plot of Figure \ref{fig: Numerical solutions.} where it halts the growth of $N_\infty$.

\section{Electromagnetic signatures of superradiant axions}
\label{sec: Electromagnetic signatures of superradiant axion clouds}

\begin{figure}[htbp]
\centering
\includegraphics[width=15cm]{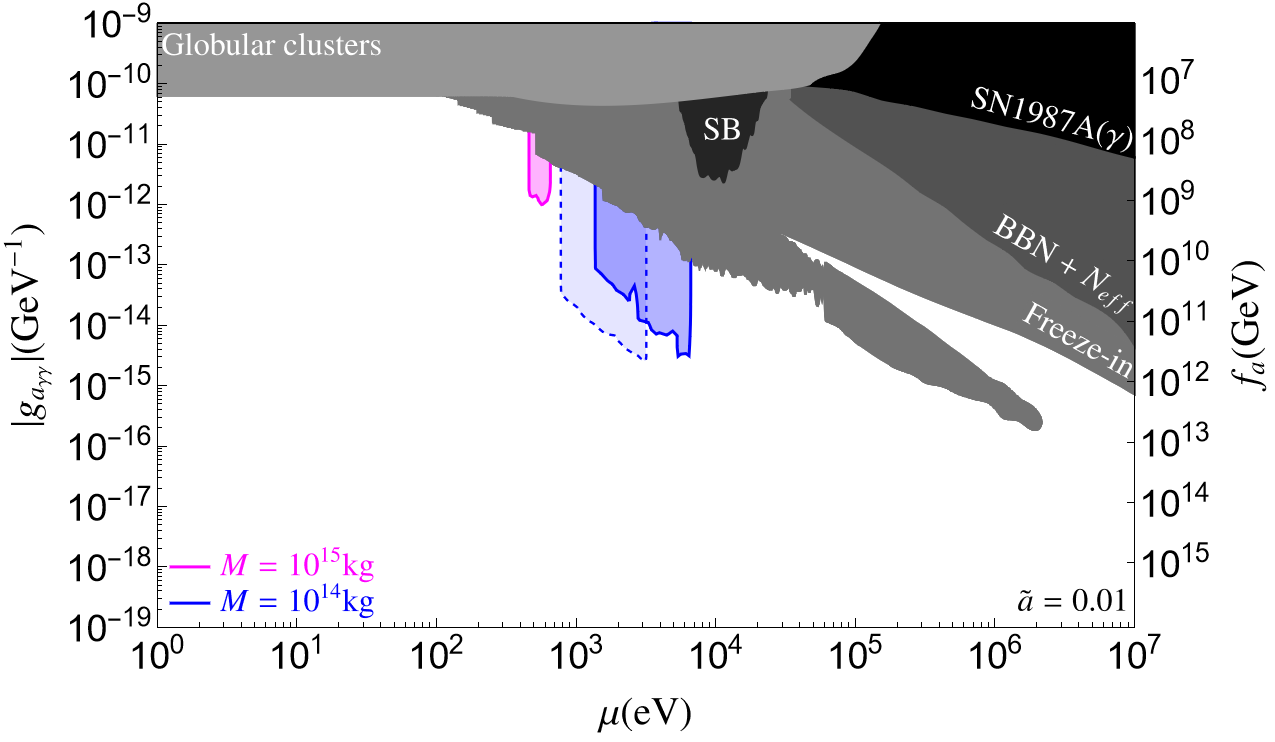}
\includegraphics[width=15cm]{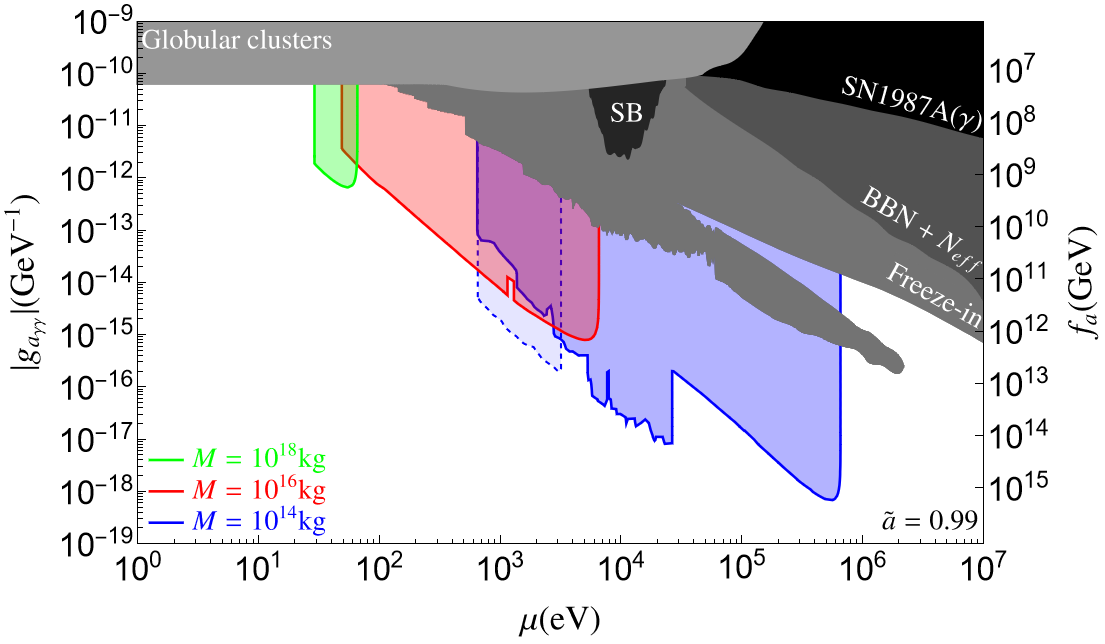}
    \caption{\small Constraints on the axion-photon coupling from PBH superradiance, assuming $100\%$ of dark matter in $10^{14}$ kg (blue), $10^{16}$ kg (red), $10^{18}$ kg (green) PBHs. The top plot is for PBHs with spin $\tilde{a}=0.01$ and the bottom plot for $\Tilde{a}=0.99$. For comparison, we show in gray/black existing constraints on axions from globular clusters, SN1987A($\gamma$), solar axions, and freeze-in production at or below BBN temperatures \cite{Ayala:2014pea,DeRocco:2022jyq,Jaeckel:2017tud,Depta:2020wmr,Langhoff:2022bij}. The predicted exclusion potential of the future ATHENA telescope is shown with a dashed blue line.}
    \label{fig: Excluded regions.}
\end{figure}

We have studied in the previous sections how a population of spinning PBHs may lead to a significant axion abundance. If these axions couple to photons they can provide interesting electromagnetic (EM) signatures of this co-existence of PBH and relatively heavy axions. In this section we will show that in a large range (see Figure \ref{fig: Excluded regions.}) of axion and PBH masses, these signatures surpass existing observational data, mainly galactic and extragalactic X-ray and gamma-ray fluxes, and that upcoming X-ray telescopes may further strengthen these constraints. 

To perform the analysis of these EM signals, we need to keep track of the total number of axions produced per PBH, both inside and outside the cloud, and their photon emission. A relevant quantity is the fraction of the dark matter in superradiant axions, given by 
\begin{eqnarray} \label{eq: fraction of axion DM}
    r =  N_\text{tot}(t_\text{uni}) \frac{\mu}{M}
\end{eqnarray}
where $M/\mu$ is the would be number of axions per PBH if axions were all the dark matter and $N_\text{tot}(t_\text{uni})\simeq N_2+N_\infty$ is the number of axions produced per PBH in each case, with $N_2$ and $N_\infty$ given  in Section \ref{sec: Clouddynamics} for the different regimes.

We will separate the discussion in two parts. First, we discuss the extragalactic emission, which is common to all cases and that originates from the extragalactic density of PBHs. 
Then, we look at situations where the ``ionized'' axions are bounded to the host galaxy.

\subsection{Extragalactic emission}

In the analysis of the extragalactic axions we use two different types of constraints. 

For axion masses between $20.4$ eV (twice the energy of the Lyman-$\alpha$ line) and a few keV, the leading source of extragalactic constraints on the axion-photon coupling are CMB spectral distortions. We use the COBE/FIRAS bounds on the rate of dark matter decays into photons derived in \cite{Bolliet:2020ofj} but re-scaled by the factor $r$ in eq. \ref{eq: fraction of axion DM}, for each value of the axion mass, to take into account the fact that the axions produced by PBH superradiance are only a small fraction of the dark matter abundance, which we assume is fully accounted for by the PBHs. 
The aforementioned constraints take into account the effects of photon injection throughout the cosmic history. We expect that for quasi-stable particles, with lifetimes greater than the age of the universe, the largest CMB spectral distortions are  generated at the latest times, when the relative energy injection into photons $\Delta \rho_\gamma/\rho_\gamma$ is maximal due to the relative enhancement of dark matter over radiation. Therefore, we expect that these constraints are also applicable in our scenario where photons are only produced at low redshifts after the superradiant production of axions takes place. However, we caution the reader that a more accurate translation of the bounds of \cite{Bolliet:2020ofj} may require a more detailed analysis.

For larger masses we estimate the background flux originated from the extragalactic axions and impose that it should be smaller than the observed one. 
We assume that the distribution of both the PBHs and the free axions is approximately isotropic. The emission rate per unit volume  
at a time $t$ is given by \cite{Masso:1997ru,Carr:2009jm},
\begin{equation}
\label{eq: photon number density}
    \frac{dn_\gamma}{dt}(E_\gamma(t),t)\simeq n_\text{PBH}(t) E_\gamma(t) \frac{d^2N_\gamma}{dE_\gamma dt}(E_\gamma(t),t)
\end{equation}
where $n_\text{PBH}(t)= \Omega_{\text{PBH}}^0 \rho_0 (1+z)^3/M$ is the PBH number density and $\Omega_\text{PBH}\simeq 0.24$ the present PBH abundance, that we fix to the dark matter abundance, and $\rho_0\simeq  8.4 \times 10^{-33}$ kg cm$^{-3}$ the critical density today \cite{Planck:2018vyg}. The redshift ($z$) factors account for the dilution due to the expansion of the universe. We have also approximated the photon emission rate as 
 $dN_\gamma/dt\simeq E_\gamma \,d^2N_\gamma/(dE_\gamma dt)$ \cite{Carr:2009jm} where $E_\gamma$ is the energy of the emitted photon. Since both the axions in the cloud and those that are ionized are non-relativistic, we may approximate the emission spectrum per PBH, $d^2N_\gamma/(dE_\gamma dt)$, as monochromatic
 \cite{Ferraz:2020zgi}
\begin{eqnarray}
\frac{dN_\gamma}{dE_\gamma dt}\simeq 2N_\text{tot} \Gamma_{a\gamma\gamma} \, \delta \left(E_\gamma(t)-\mu/2 \right)\, .
\end{eqnarray}
where $N_\text{tot}$ is the total number of produced axions in eq. \ref{eq: Ntot} at a given time.
To obtain the photon flux, $I\simeq  \,n_\gamma(t_\text{uni})/(4\pi)$ \footnote{We remark that we are using natural units, otherwise a factor $c$, the speed of light, should appear in the relation between $I$ and $n_\gamma$.}, we integrate eq. \ref{eq: photon number density} in time until today and find
 \begin{equation}
     I(E)= \frac{3}{4\pi}\Omega_{PBH}^0\, \rho_0 \,t_\text{uni} \,\Gamma_{a\gamma\gamma}\frac{N_\text{tot}(t_e)}{M} \left(\frac{E}{\mu/2}\right)^{3/2}
 \end{equation}
where $t_e$ is the time of emission of the photon that arrived today with energy $E$ and that can be obtained through the redshift relation $1+z_e(t_e)=\mu/(2E)$. We have used that most photons are emitted during the matter-dominated epoch given the timescales involved in the dynamics of the superradiant clouds (and neglected the current era of dark energy domination).

The flux at energies $E<\mu/2$ corresponds to the photons emitted before today. As we have mentioned at the end of the previous section, the total number of axions never decreases over time; the decays into photons at most halt its growth. Therefore, most of the EM flux will be due to the non-relativistic axions that are around ``today'' and so to compare with data we simply evaluate the flux at $E=\mu/2$ for each value of the parameters ($\mu,M,g_{a\gamma\gamma}$). 
We define the excluded regions of parameters as those where the EM flux is larger than the double power law fit to the X-ray and gamma-ray background data used in \cite{Ballesteros:2019exr}.

\subsection{Galactic emission}
The axions within the cloud and those that are ionized with sufficiently small velocities remain bound to the host galaxy. Their decay into photons contributes to the galactic background fluxes where observational constraints are typically stronger. 
 We assume that their density tracks the dark matter profile in the galaxy but, similarly to the COBE/FIRAS bounds derived above, with an amplitude that is suppressed by the fraction $r$ in eq. \ref{eq: fraction of axion DM} so that we can rescale existing constraints on the rate of dark matter decays into photons from the Andromeda \cite{Ng:2019gch}, Milky Way \cite{Laha:2020ivk,Foster:2021ngm} and Leo-T \cite{Wadekar:2021qae} galaxies. Finally,
 we study the prospects of detection with future ATHENA X-ray telescope using the forecasted bounds on dark matter decays into photons from Msec observations of the Segue 1 dwarf spheroidal galaxy \cite{Neronov:2015kca}. 

\subsection{Discussion of the constraints}
The resulting constraints in the $g_{a\gamma \gamma}$ vs $\mu$ plane are shown in Figure \ref{fig: Excluded regions.} for slowly rotating ($\tilde{a}=0.01$, top) and extremal ($\tilde{a}=0.99$, bottom) PBHs, and for different values of their mass (in different colors). The constraints are stronger in the case of near-extremal black holes because the spin-down of the black hole leads to a larger axion abundance.

For each value of the PBH mass the boundary of the constrained region is delimited on the right (large $\alpha$) by the superradiance condition and on the left (small $\alpha$) by the fact that the photon flux decreases for smaller axion masses. 
On the other hand, by increasing $f_a$ at $\mu$ we move from the \textit{no spin-down} regime into the \textit{incomplete spin-down} and eventually into the \textit{complete spin-down} region. In the \textit{incomplete spin-down region}, a sizeable portion of the PBH spin is already extracted hence $N_\infty$ is almost maximal. Therefore, increasing $f_a$ further will only decrease the flux, and that is even more so in the \textit{complete spin-down} where the flux decays as $f_a^{-2}$.

In the near-extremal case, there is a sudden change in the constrained region, around $\mu\simeq 30$ keV for $M=10^{14}$ kg. This change originates from the fact that for larger masses (large $\alpha$) the emitted axions have escape velocities larger than $10^{-3}$ and so only the extragalactic bounds can be applied, whereas for smaller masses the Milky Way and Andromeda constraints on the fluxes can be used. The same logic explains why the forecasted sensitivity of ATHENA stops at a few keV; the analysis of \cite{Neronov:2015kca}, used in this work, relies on an observation of the Segue 1 galaxy where the escape velocity is of order $10^{-4}$.

Finally, we compare the constraints obtained in this work with existing constraints on the axion-photon coupling that do not rely on an initial thermal or dark matter axion abundance and that originate from: globular clusters \cite{Ayala:2014pea}, axions accumulating in the solar basin \cite{DeRocco:2022jyq}, supernova 1987A \cite{Jaeckel:2017tud}, and freeze-in production at or below BBN temperatures \cite{Depta:2020wmr,Langhoff:2022bij}. We have extracted this data from the repository \cite{AxionLimits}. Our constaints are stronger than the existing constraints for a wide range of parameters, particularly in the case of near-extremal PBHs for which superradiant axion production is most efficient. X-ray telescopes such as ATHENA will be able to probe this scenario even further.

\section{Conclusion}
\label{sec: Conclusion}

PBHs with asteroid-like masses in the range $10^{14}-10^{18}$ kg can account for all the dark mater but are hard to probe due to their minuscule size. In this work, we showed that the co-existence of an axion with a $0.1-10^6$ eV mass can dramatically change this picture.

The mechanism underlying our study is black hole superradiance: the draining of the black hole's spin into a densely packed bosonic cloud of a field with mass below the black hole's inverse radius.
We considered two benchmark values for the PBH spin parameter, slow-spin $\tilde{a}=0.01$ and near-extremal $\tilde{a}\simeq 0.99$, which correspond to PBHs born in the radiation era or an early matter era, respectively; and restricted the axion and PBH masses to ranges where superradiant clouds form within the age of the universe (c.f. Figure \ref{fig: Parameter space for the ALP.}).

Building upon the work of \cite{Baryakhtar:2020gao}, we then studied the cloud's evolution in the presence of the characteristic axionic self-interactions, the main effect of which is to trigger a non-linear mixing between the two dominant superradiant states, $\left|211\right>$ and $\left|322\right>$, alongside with axion reabsorption by the black hole and emission to infinity of axions ionized from the superradiant cloud. We studied the dynamics of the system by solving numerically and analytically a coupled set of Boltzmann equations for the number of particles in the leading superradiant levels alongside the evolution equations for the PBH mass and spin.

We identified four main regimes in the dynamics (c.f. Table \ref{table with regimes}), that are characterized by different ranges of the axion decay constant $f_a$, and provided an accurate analytical description in Section \ref{sec: Boltzmann equations}.  
In all regimes the cloud starts growing exponentially due to the superradiant instability, which occurs on cosmological timescales even for slowly spinning PBHs. However, when self-interactions are strong $f_a<f_a^{\text{spin-}\downarrow}$, the cloud quickly reaches an equilibrium state where the occupation numbers of the leading superradiant levels are approximately constant. This is sustained by the superradiant instability that continuously drains spin from the PBH but at a slower rate. We labelled this regime \textit{no spin-down}.

For $f_a^{\text{spin-}\downarrow}<f_a<f_a^\text{no-spin}$, self-interactions become less efficient and the quenched equilibrium is reached later, when the cloud is denser. To keep this denser equilibrium, the black hole has to lose spin at a much faster rate and so, after some time, it eventually loses a significant portion of its initial spin. At this point, the system enters a new adiabatic regime where both the spin and the equilibrium numbers decrease slowly. This is the \textit{incomplete spin-down} regime.
The \textit{complete spin-down} regime occurs for even weaker self-interactions, $f_a^\text{no-spin}<f_a<f_a^\text{no-eq}$, when the adiabatic regime is able to drain so much spin that the superradiance condition is no longer satisfied and the black hole can no longer refill the cloud. From then on, the fate of the cloud is fully determined by the self-interactions that continue dissipating axions to infinity and back into the black hole (through non-superradiant bound states). 
Finally, for $f_a>f_a^\text{no-eq}$, self-interactions are so weak that the cloud reaches its maximal occupation number, by shutting down the (leading) superradiant instability before any quenched equilibrium is reached. This \textit{fast spin-down} regime is equivalent in its dynamics to the vanilla case of superradiance of a non-interacting massive bosonic field.

In the last three regimes mentioned above (\textit{incomplete, complete} and \textit{fast spin-down}) the PBH loses a significant part of its spin before the present day. This energy is lost into axions that either escaped to infinity (\textit{incomplete} and \textit{complete spin-down}) or are still mostly within the cloud (\textit{fast spin-down}). Therefore, the final axion abundance can be very large, in particular in the case of highly-spinning PBHs, and this begged the question, what happens to these axions? Can they give observational signatures of this scenario?  

We addressed this in Section \ref{sec: Electromagnetic signatures of superradiant axion clouds} where we assumed that the produced axions decay into photons, with a rate directly related to the strength of their self-interactions (i.e.~we set the electromagnetic anomaly coefficient in eq. \ref{eq: Axion Lagrangian} to one), and explored the resulting electromagnetic signals assuming a monochromatic PBH spectrum accounting for all dark matter.  We identified two main signatures: \textit{i)} an extragalactic photon flux that results from the decay of axions produced by the  cosmological PBH population; and \textit{ii)} galactic fluxes from the axions that are ionized from the cloud with a velocity below the host galaxy's escape velocity. We imposed constraints from galactic and extragalactic (X-ray and gamma-ray) background data, and existing constraints on the rate of dark matter decay into photons from COBE/FIRAS and the Leo-T dwarf galaxy (see Section \ref{sec: Electromagnetic signatures of superradiant axion clouds} for more details), all properly rescaled to account for the axion abundance, to derive bounds on the axion mass and decay constant $f_a$. Finally, we have also shown how future X-ray telescopes such as ATHENA will be able to further probe this scenario. These results are summarized in Figure \ref{fig: Excluded regions.} and show that the assumption of PBH dark matter excludes the existence of $20-10^6$ eV axions coupled to photons for a broad range of  $f_a$ values, depending on the PBH mass and spin. 

There are a few interesting aspects that would be interesting to further explore in the future. A sub-leading effect that we have discarded in our analysis is the emission of gravitational waves by the axion clouds. Although this has no significant impact on the dynamics, it may potentially provide additional observational signals. Given that typically only a very small fraction of the PBH mass is transferred to the superradiant clouds, observing individual clouds is virtually impossible, but the large PBH abundance may potentially lead to an observable stochastic background, which we plan to investigate in the future.

In our analysis we have followed the results obtained in \cite{Baryakhtar:2020gao} in what concerns the leading effects of axion self-interactions. Since the dimensionless mass coupling $\alpha<1/2$ in all cases considered, we expect a non-relativistic approach to be at least a reasonable approximation, although one may expect considerable corrections for the larger $\alpha$ values attained in the near-extremal regime such as the Bosenova regimes observed in \cite{Yoshino:2012kn, Yoshino:2015nsa, Omiya:2022mwv,Omiya:2022gwu}.

In addition, this analysis also neglects higher-order interactions (from the axion potential). Even though the equilibrium number of axions within the cloud is always sufficiently small that the field value $a\lesssim f_a$, it would be interesting to take a closer look into the significance of higher-dimensional interactions leading to multiple axion scattering processes, as well as into how non-linearities affect the cloud's density profile and growth rate \cite{Omiya:2022mwv}.

Our results should therefore be viewed in the light of the approximations employed and motivate a further exploration of these issues. Nevertheless, an important take-home message of this work is that self-interactions only change the timescale at which the PBH lose most of its spin. Given enough time, the total number of axions produced (whether they remain or not bound to the black hole) approaches the maximum number produced in the absence of self-interactions, up to $\mathcal{O}(1)$ factors. In our analysis, we have found that this occurs within the universe's age in most cases except for very low values of the axion decay constant.

We hope that our results motivate further studies of PBH superradiance and of its relevance for fundamental particle physics and cosmology.

\section*{Acknowledgements}

RZF is supported by the Direcci\'o General de Recerca del Departament d’Empresa i Coneixement (DGR) and by the EC through the program Marie Sk\l odowska-Curie COFUND (GA 801370)-Beatriu de Pinos. NPB is supported by Centro de Física da Universidade de Coimbra (CFisUC) and by Fundação para a Ciência e a Tecnologia, I.P./MCTES through national funds (PIDDAC). This work was supported by the CFisUC project No.~UID/FIS/04564/2020 and by the FCT-CERN grant No.~CERN/FIS-PAR/0027/2021.

\appendix

\bibliography{mybibfile}

\end{document}